# Pressure safety approach for PIP-II cryogenic distribution system and cryomodules

William Soyars[1], Tomasz Banaszkiewicz[2], Ram Dhuley[1]
[1]Fermi National Accelerator Laboratory, Batavia, Illinois 60510, USA
[2]Wroclaw University of Science and Technology, Poland

Email: soyars@fnal.gov

**Abstract**: The Proton Improvement Plan-II (PIP-II) is a superconducting linear accelerator being built at Fermilab that will provide 800 MeV proton beam for neutrino production. The linac consists of a total of twenty-three (23) cryomodules of five (5) different types.  Cooling is required at 2K, 5K and 40K. The Cryogenic Distribution System (CDS) consists of a Distribution Valve Box, ~285 m of cryogenic transfer line, modular Bayonet Cans to interface with cryomodules, and a Turnaround Can. The cryogenic system must provide protection from over-pressure by sizing pressure relief devices for all volumes and process line circuits.  The cryomodule cavity circuits have dual pressure ratings, 4.1 bara when cold and 2.05 bara when warm (T>80K).  Worst case relieving cases will be identified. The methods for determining heat flux will be presented. For the relieving occurring in the linac tunnel, flow must vent to outside to avoid an oxygen deficiency hazard.  Also, we will present vacuum vessel relief sizing to protect the cryogenic distribution system vacuum shells from over pressure during an internal line rupture.  The project is funded by US DOE Offices of Science, High Energy Physics.

## 1. Introduction

The Proton Improvement Plan – II (PIP-II) is a superconducting linear accelerator being built at Fermilab that will provide 800 MeV proton beams for neutrino production. The cryogenic distribution system (CDS) must deliver 2K, 5K, and 40K cooling from the cryogenic plant to the linac cyromodules (CM). The linac contains five unique cryomodules. The CDS is segmented to provide each cryomodule with its own dismountable connect for each process line.  The Cryogenic Distribution System and cryomodules are schematically shown in Figure 1.  Thermodynamic design aspects have been investigated previously [1].

The CDS, from the Distribution Valve Box at the interface with the Cryogenic Plant, through twenty-five Bayonet Cans (BC), to the Turn-Around Can at the end of the linac, is approximately 285 m long. The CDS process circuits with size and nominal operating temperatures are: 4.5K Supply- DN 50 at 4.5K, 2K Return- DN 250 at 3.8K, Low Temperature Thermal Shield (LTTS) Return- DN50 at 9K, High Temperature Thermal Shield Supply (HTTS-S) – DN50 at 40K and Return (HTTS-R)-DN 50 at 80K, Cooldown Return (CD) – DN 80 at 80K.

All piping and vessels must be protected from overpressure for safe operation. The relief device locations are indicated in Figure 2. The pressure safety requirements established for the project and the technical approach will be useful for comparing to other systems.

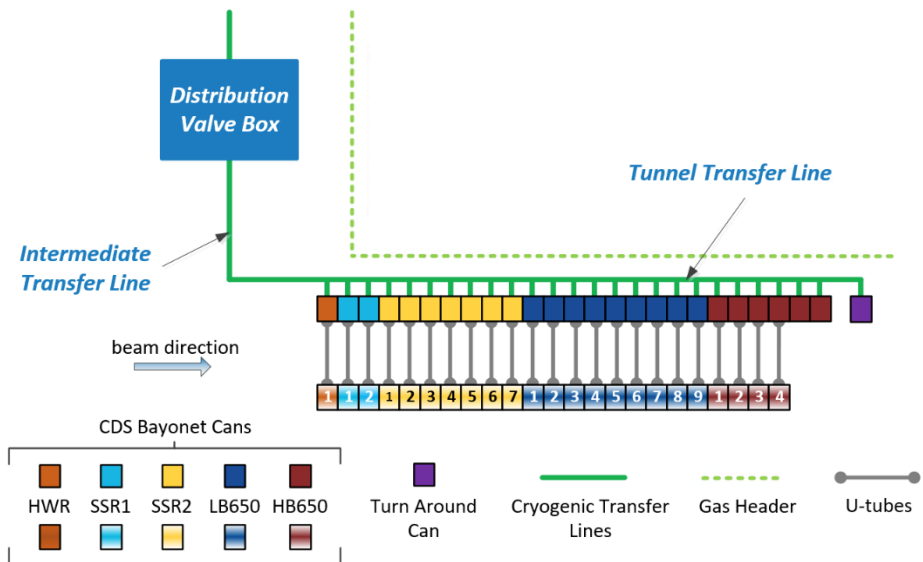

**Figure 1.** PIPII cryogenic distribution system and cryomodules schematic

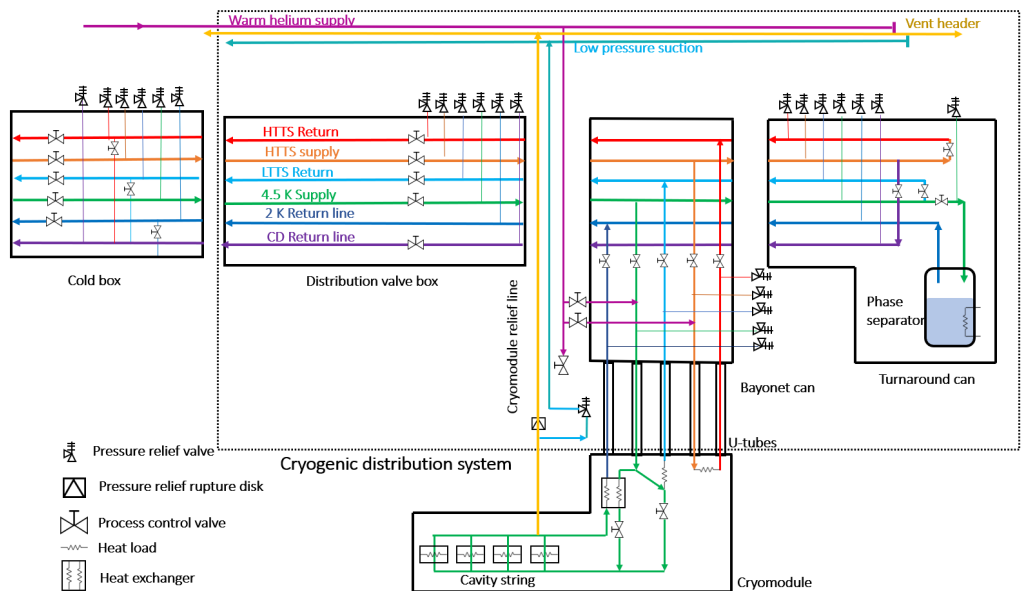

**Figure 2.** PIPII cryogenic distribution system and cryomodules relief device locations

## 2. Requirements

To meet project safety and quality requirements, the system will comply with ASME Boiler and Pressure Vessel Code and ASME B31.3 Process Piping Code. The CDS relieving is available at each end of the CDS. CM positive pressure relieving is available at the BC, connected by the u-tubes. CM 2K relief devices will be directly mounted on each CM. For the 2K circuit of the CMs, a dual pressure rating is given [2]. This utilizes the higher allowable stress of the niobium cavities when cold, <80K.

Any helium process line relieving into the linac tunnel space is not allowed. This enhances the Oxygen Deficiency Hazard safety for personnel in the tunnel. All reclosable safety relief valves will exhaust into the Low Pressure Return to compressor suction. For the CM 2K Return SVs which nominally operates with a sub-atmospheric pressure at its inlet, this is very important for system purity so that any leakage through the relief poppet will not have air contamination. And for this positive pressure SVs, this will save on helium inventory losses to recover any leakage from positive pressure process circuits.

## 3. Calculation Methods and assumptions

Sources of overpressure must be identified. For cryogenic systems, a significant source of overpressure is a sudden loss of insulating vacuum. Furthermore, for cryomodules, there is an additional vacuum failure mode from sudden loss of beam tube vacuum. Other potential sources of overpressure must also be considered. For some warmer circuits, with air condensation being less of a problem, cryogenic plant supply with return flow isolated will be investigated.

Temperature conditions at cryogenic relieving are set according to standard methods [3,4]. Pressure conditions are set by Code allowable overpressure. Take this as 21% above Maxiumum Allowable Working Pressure (MAWP) by considering loss-of-vacuum (LOV) as an "unexpected source of external heat" [5]. This allows for 21% above MAWP for "vessels exposed to fire or other unexpected sources of external heat." According to ISO 4126 Terms and Definitions for Pressure (section 3.2) the maximum allowable overpressure "is established by applicable code for and fire conditions". Calculations are performed in accordance with the EN ISO 4126-1 method for gas, and also with direct integration method which can lead to different results at cryogenic conditions.

### 3.1. Heating during loss of vacuum

One straightforward method to is to assume a constant, maximum heat flux along the length of the CDS or CM. This will apply known, experimental heat fluxes [6,7]. One term, for thermal radiation shield internal side without MLI, natural convection calculations used.

A second method commonly used [6,7] is to limit the available heat input to the energy available energy deposition of air through the worst case, feasible orifice in the vacuum vessel. This often is the most plausible representation for a long transfer line scenario. See Table 3 for worst-case air inleak orifice assumption. When there are multiple process lines available for air condensation, this total energy needs to be distributed among the lines. The cold surface absorbs heat proportionately to its surface area as a percentage of the total cold surface area and temperature dependent heat flux product.

### 3.2. Metal heating during loss of vacuum

For further refinement, one can consider that not all of the air ingress energy heats the helium but that some of air ingress heating goes to heating the metal of the CDS. This will lower the heat load to helium which the relief system must address. Consider for the given mass of piping, the heat capacity energy needed to raise metal temperature from process fluid temperatures to temperature where air freezing/condensing ends, taken to be 70K. To convert to heating rate, need time constant to apply for the duration of the heating period. Use same time constant as previously used for LCLSII transfer line which has similar geometry [8], 70 sec. This is based on experimental results from DESY/XFEL Cryomodule crash tests [ref] for air ingress choked flow, with active cryo-pumping, using same size opening to vacuum as considered here, DN80.

## 4. Cryogenic distribution system protection from overpressure

Geometry is taken from the preliminary design solid model. Each process line has two primary reliefs: one at the Distribution Valve Box (DVB) and one at the Turnaround Can (TC). Thus, CDS relieving is available at each end of the CDS.

### 4.1. Heat input from loss of insulating vacuum

Assume air leaks in through open DN80 vacuum evacuation port. Applying approach in 3.1, this orifice size limits energy and is used to define loss-of-insulating-vacuum (LOIV) heat flux. See Table 1 for results.

### 4.2. Overpressure flow rate from cryoplant

Consider that the cryogenic plant is operating at full capacity for a given circuit, then its return valve is closed, making the cryoplant a source of oversupply. This is conservative because realistically the plant cannot output full capacity at its maximum design pressure. Nevertheless, this is basis for design for the HTTS Supply and the CD Return, which have very small requirements from LOIV. See Table 2.

*4.3. Relief sizing requirements*

Apply heat transfer to He as shown in Table 1. Then relief requirements are calculated in Table 2.

**Table 1.** Heat load to CDS circuits during LOIV

| Circuit | Peak Heat Flux | Dia. | Heat load per Length, from experimental flux | Total Heat load distributed to He, from constant flux | Heat transfer to He | | Heat transfer to metal | | TOTAL heat transfer |
|---|---|---|---|---|---|---|---|---|---|
| | [kW/m²] | [cm] | [W/cm] | [kW] | [kW] | [%] | [kW] | [%] | [kW] |
| 2k Return | 6 | 27.3 | 51.46 | 42% | **232** | 35 | 80 | 12 | 312 |
| 4.5k Supply | 6 | 6.03 | 11.37 | 9% | **51** | 8 | 11 | 2 | 63 |
| LTTS | 6 | 6.03 | 11.37 | 9% | **51** | 8 | 11 | 2 | 63 |
| HTTS Supply | 0.23 | 6.03 | 0.44 | 0% | **2** | 0 | 9 | 1 | 11 |
| HTTS Ret & Shield, insulated | 0.23 | 62 | 4.48 | 4% | **20** | 3 | 0 | 0 | 20 |
| HTTS Shield, uninsulated | 2.4 | 56 | 42.22 | 35% | **191** | 29 | 0 | 0 | 191 |
| CD | 0.23 | 8.89 | 0.64 | 1% | **3** | 0 | 0 | 0 | 3 |
| **SUM** | - | - | **127** | **100%** | **551** | **83** | **112** | **17** | **663** |

**Table 2.** Helium relieving requirements for CDS

| Circuit | Set Press. [kPa] | Relieving Temp. [K] | Worst case condition | Heat load [kW] | Helium flow requirement, each relief [kg/s] | SV size, DN |
|---|---|---|---|---|---|---|
| **2k Return** | 410 | 7.0 | LOIV | 232 | 4.78 | 100 x 150, two |
| **4.5k Supply** | 2000 | 12.7 | LOIV | 51 | 0.34 | 20 x 25, two |
| **LTTS Return** | 1000 | 9.7 | LOIV | 51 | 0.57 | 20 x 25, two |
| **HTTS Supply** | 2400 | 40. | Cryoplant oversupply | Not applicable | 0.055 | 20 x 25, two |
| **HTTS Return & shield** | 2400 | 80. | LOIV | 211 | 0.24 | 20 x 25, two |
| **CD Return** | 1000 | 80. | Cryoplant oversupply | Not applicable | 0.10 | 20 x 25, two |

## 5. Cryomodules protection from overpressure

This analysis applies to the five different CMs: HWR, SSR1, SSR2, LB650 and HB650 CM. CM positive pressure relieving is available at the BC, connected by the u-tubes. CM 2K relief devices will be directly mounted on each CM. For the 2K circuits, these require further analysis due to their dual pressure ratings; furthermore, these have a unique loss of vacuum failure mode due to the LOBV scenario. None of these LOV scenarios will take credit for metal warming.

### 5.1. Non-2K circuit pressure relieving from loss of insulating vacuum

Geometry is taken from solid models of the five different styles. For the positive pressure circuits, the heat load will be simply and conservatively estimated as the peak constant heat flux values (shown in Table 1) applied to its surface area. The cryomodules have a 2K-to-4K subcooling heat exchanger, and an approximation is made for heat exchanger surface area, with the sides and lengths treated as a cube. All heat flux to heat exchanger is assumed to go to the 4.5K supply. HTTS geometry is looked at separately, but then the heat load is divided between the 80K and 40K circuits, since this piping is common volume to both reliefs and acts over gradient of temperature. Results shown in Table 3.

### 5.2. Rupture Disk sizing for sub-atmospheric 2K circuit cold MAWP protection

The goal is to protect the CM SRF cavities vessel from overpressure for cold conditions where the MAWP is 410 kPa. Use HB650 CM cavity geometry, the worst case. Two LOV scenarios will be considered. For loss-of-insulating-vacuum (LOIV), assume air leaks in through open DN100 vacuum evacuation port. For loss-of- beam tube-vacuum (LOBV), assume air leaks in through 60 mm diameter coupler port. Applying approach in 3.1, neither orifice size limits energy; thus, a constant heat flux over the surface area is used to define LOV heat flux.

Apply peak heat flux values to surface area. When comparing LOIV and LOBV case, one sees LOBV is the worst case, and will be used to define the requirements. See results in Table 3. This leads to DN100 RD, a sizing result previously arrived at to protect CMs while undergoing testing and is in service on our CM test stands.

**Table 3.** Helium relieving requirements for Cryomodules

| Circuit | Set Press. [kPa] | Relieving Temp. [K] | Worst case condition | Heat load [kW] | Helium flow requirement, each relief [kg/s] | SV size, DN |
|---|---|---|---|---|---|---|
| **4.5k Supply** | 2000 | 12.7 | LOIV | 3.23 | 0.047 | 15 |
| **LTTS Return** | 2000 | 12.7 | LOIV | 10.6 | 0.16 | 15 |
| **HTTS Supply** | 2400 | 40. | LOIV | 43.9 half of HTTS total | 0.20 | 15 |
| **HTTS Return & shield** | 2400 | 80. | LOIV | 43.9 half of HTTS total | 0.10 | 15 |
| **2k volume-cold** | 410 | 7.0 | LOBV | 216 | 8.9 | 100 |
| **2k volume-cold** | 275 | 6.0 | LOBV | 166 | 8.5 | 100 |
| **2k volume-warm** | 205 | 300 | Oversupply with return closed | Not applicable | 0.27 | 80 x 100 |

*5.3. Safety valve sizing for 2K circuit warm MAWP protection*
While the CMs are warm, >80K, they can see pressure above allowable due to oversupply from the cryogenic transfer line. Consider HB650 CM, which is the largest and have the largest cooldown and JT supply valves. For worst case flow conditions, consider that the CDS TL is cold at nominal conditions. The CM is warm, but 4.5K He supply valve is mistakenly set full open. Assume both CM JT and CD valves with known sizing also open. Assume valve closed, isolating the gas return path. Results are shown in Table 3.

## 6. Venting the relief exhaust to atmosphere
A final function of the CDS is to provide for warm gas return to cryoplant compressor suction during purification, cooldown, and warm up operations. This Low Pressure (LP) Return piping will be DN200 pipe and is rated at 140 kPa. The Design Pressure is consistent with other systems at Fermilab.

*6.1. LP Return for CDS and CM reclosable relief devices*
The CDS requires for safety valves to exhaust into the LP Return header. This will be accomplished with SVs manifolded into a collection header at TC and DVB, which is then flanged into the LP Return. A key constraint on this is to ensure backpressure is low enough to not impact the 2K Return reliefs, set at 410 kPa. This necessitates that the 2K Return reliefs are a balance bellows style, which can accommodate some backpressure variation, up to about 50% of the gauge set pressure. This guided the set pressure choice of the LP Return header at 140 kPa, to allow for pressure drop up to the outdoor low pressure relief. There is relief to atm at either end of the LP Return header. At the DVB, it's a short distance from inside the CP building through a hole in the wall. At the TC, its more complicated because of limited penetrations from the tunnel up to the surface. Calculations show we want to expand the LP from 8 to 10 to meet backpressure requirements.

There is relief to atmosphere at either end of the LP Return header. At the DVB, it's a short distance from the DVB inside the CP building through a hole in the wall. At the TC, its more complicated because of limited penetrations from the tunnel up to the surface. The closest available one is about 30 m away. Calculations show for this header run up to the outdoor relief want to expand the LP from DN200 to DN250 to meet backpressure requirements. The plan is to use Fermilab designed and fabricated parallel plate spring loaded reliefs, with known performance from previous testing and operations [9].

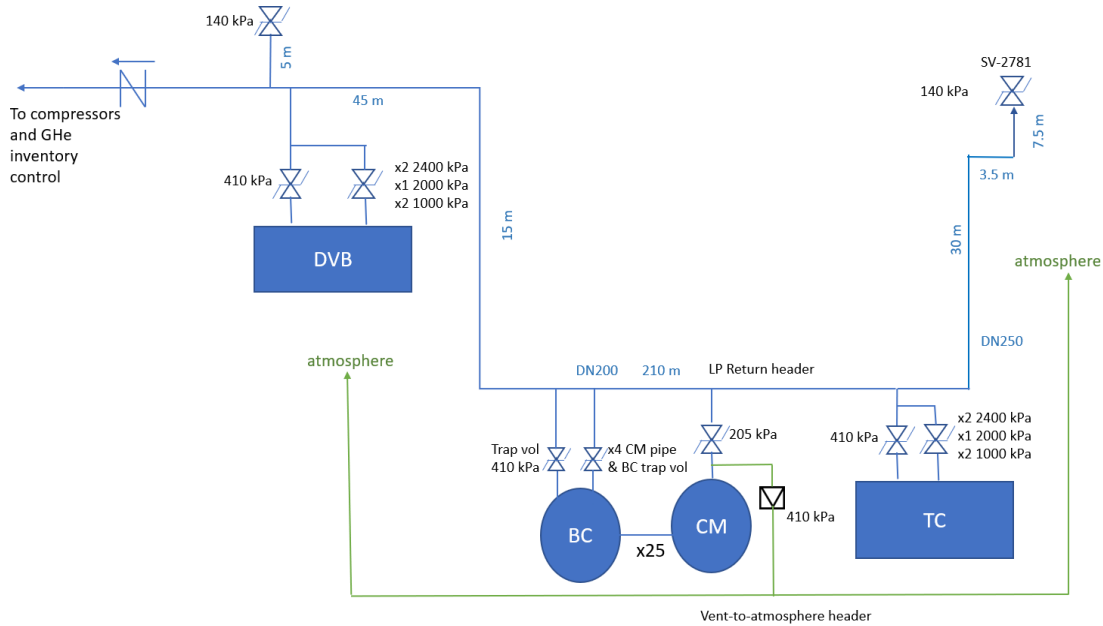

**Figure 3.** Schematic of LP Return and venting to atmosphere

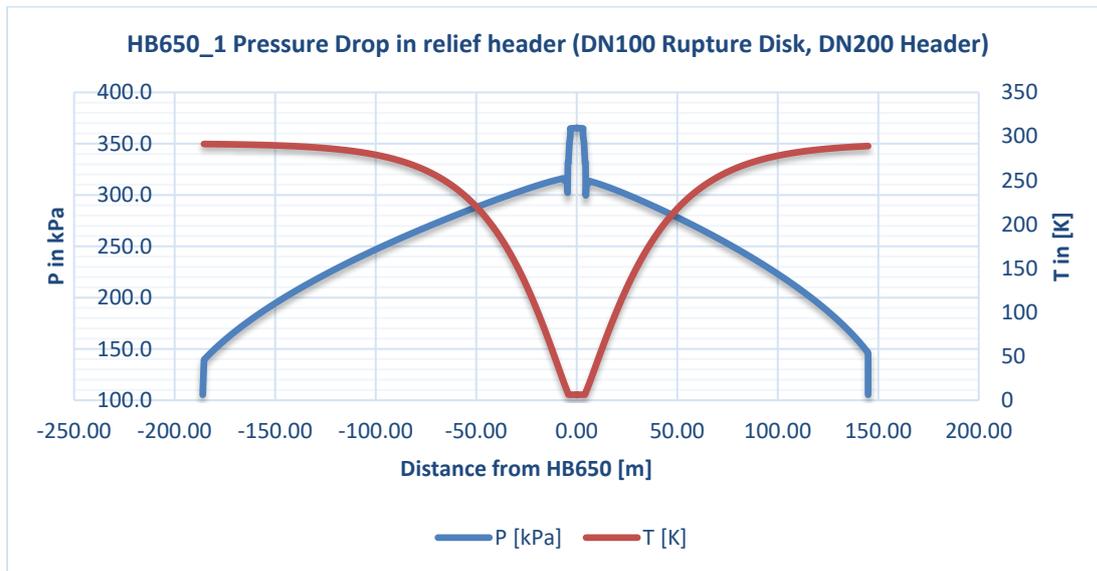

**Figure 4.** vent to atm DP

*6.2. Tunnel vent-to-atmosphere header for CM 2K RD*

Requirements call for RD flow to vent to outside, not into the tunnel. This will utilize a dedicated line open to atmosphere at the surface. Analyse the first HB650 in the tunnel, which is the worst case as it at the midway location between the two vent chimneys to atmosphere. This location is 150 m to the US linac penetration and 131 m to the DS linac tunnel penetration

Calculations are made using internal HB650 CM geometry. This includes flow from eight cavities, collecting into common DN100 pipe to the RD. Conservatively assume a check valve present; this is conservative since production CMs are not expected to include one. After exiting the vacuum jacket space from the CM, temperature rise is calculated using natural convection heat transfer. For vent piping thermal contraction, an axial bellows is assumed at each of twenty-five BC, and these are considered in the flow resistance analysis.

The goal is to ensure that open RD opening, the cavities do not exceed 121% of their cold MAWP, that is 496 kPa. A DN200 vent-to-atmosphere header accomplishes this. See Figure 4. This size matches that of the LP Return header, which has advantage of commonality in the design.

## 7. Vacuum vessel protection from overpressure

The CDS vacuum jacket is size DN700 and must be protected from overpressure. It has a MAWP of 150 kPa. There are three insulating vacuum space volumes. The lengths are 71 m for surface and vertical penetration, then 104 m and 106 m in the tunnel. Note, CM vacuum vessel protection will not be discussed here.

Consider the spontaneous rupture of one internal line as the source of overpressure. For the most conservative approach, the analysis assumes that the line ruptures along the entire circumference of the pipe. The following are calculated: volume of vacuum jacket space, fixed mass of He in the pipe which is suddenly spilled, new He density, temperature to reach 150 kPa, and heat inflow primarily from outside jacket (300K) and secondarily from thermal shield (80K). Capacity and sizing are done with standard methods [4, 10].

It was found that 4.5K Supply generated the worst case. It will contact warmer surfaces of the vacuum vessel (300K) and the HTTS shield (80K), absorb heat, and expand. Consider maximum cryoplant capacity of 200 g/s at 300 kPa as providing flow and use know geometry for cross-sectional area in the DN700 pipe; this is used to come up with Reynolds number for approximating heat transfer coefficient from external heating, using MacAdams equation, a basic relationship. For the three segments, see Table 5. A design choice is made for relief diameter of 70 mm, which leads each TTL-BC to have two reliefs.

**Table 5.** Vacuum relief requirements for CDS

| Ruptured pipe | Vacuum segment | Heating [kW] | Mass flow generated [kg/s] | Total $A_{SV}$ [mm$^2$] | Required number of relief valves (70 mm diameter) |
|---|---|---|---|---|---|
| 4.5K Supply | **DVB + ITL** | 1037 | 72 | 59434 | 16 |
| | **TTL volume 1, 13 BCs** | 1443 | 100 | 82680 | 22 |
| | **TTL volume 2, 14 BCs** | 1490 | 103 | 85383 | 23 |

The method above requires helium gas leaking into any one location to flow to adjacent piping for relieving. A final check was made to ensure that the fixed supports along the ID of the vacuum jacket are not serving as a restriction. The flow area through the fixed support is >> the orifice area of one RV.

## 8. Conclusions
The CDS and CM pressure safety requirements, methods, and results have been presented. Flow requirements are available for relief device sizing specifications for design. These results are for preliminary design; final selection will occur during detailed design and procurement process. Nominal room temperature piping requirements to meet system relieving needs have been established.

**Acknowledgments**
This manuscript has been authored by Fermi Research Alliance, LLC under Contract No. DE-AC02-07CH11359 with the U.S. Department of Energy, Office of Science, Office of High Energy Physics. The authors wish to thank Erik Voirin of Fermilab for his inputs and reviewing of this work. This work benefited from similar analyses by Jay Theilacker and Andrew Dalesandro of Fermilab for the LCLSII project at SLAC.